# A piecewise-linear reduced-order model of squeeze-film damping for deformable structures including large displacement effects

A.Missoffe[1], J.Juillard[1], D.Aubry[2]

[1]SUPELEC, Dpt. SSE, 3 rue Joliot-Curie, 91192 Gif-sur-Yvette, FRANCE.
E-mails: alexia.missoffe@supelec.fr, jerome.juillard@supelec.fr.

[2]LMSS-MAT, ECP, CNRS UMR 8579, Grande Voie des Vignes, 92295 Châtenay-Malabry Cedex, FRANCE. Email : denis.aubry@ecp.fr.

*Abstract-*. This paper presents a reduced-order model for the Reynolds equation for deformable structure and large displacements. It is based on the model established in [11] which is piece-wise linearized using two different methods. The advantages and drawbacks of each method are pointed out. The pull-in time of a microswitch is determined and compared to experimental and other simulation data.

I. INTRODUCTION

Correct modelling of damping is essential to capture the dynamic behaviour of a MEMS device. Our interest is squeeze-film damping which models the behaviour of a fluid in small gaps between a fixed surface and a structure moving perpendicular to this surface. The lateral dimensions of the surfaces are large compared to the gap and the system is considered isothermal. Squeeze film damping is then governed by the Reynolds equation [1]:

$$\nabla \left( \frac{G^3}{12\mu} P \nabla P \right) = \frac{\partial (GP)}{\partial t} \quad (1)$$

where $G(x,y,t)$ is the distance between the moving and the fixed surface, $P(x,y,t)$ is the pressure, $\mu$ the effective viscosity of the fluid [1]. For small excitation frequencies or amplitudes the squeezed film behaves as a linear damper. For larger amplitudes or frequencies, the gas has no time to flow away and the pressure builds up creating a stiffening effect coupled to a nonlinear damping. A complete review on this equation and its different regimes can be found in [3].

Coupling the Reynolds equation to the equation governing the mechanical behaviour of the microstructure leads to a nonlinear system of partial differential equations (PDEs), which has no analytical solution and must be simplified thanks to some assumptions. The most commonly made assumptions are the following:

- uniform displacements, i.e. $\frac{\partial G}{\partial x} = \frac{\partial G}{\partial y} = 0$

- steady-state sinusoidal excitation, i.e. $G = G_e \sin(\omega t)$ [4]

- small displacements, i.e. $G = G_0 + g$ and $g \ll G_0$, where $G_0$ is the nominal gap of the structure at rest or close to a static equilibrium [5].

- small pressure variations, i.e. $P = P_0 + p$ and $p \ll P_0$, where $P_0$ is the ambient pressure.

These hypotheses prove to be useful in a variety of applications, if only for gaining insight of nonlinear damping phenomena. However, in many cases, it is difficult to justify their use: for example, it is clear to see that none of the first three hypotheses holds when trying to estimate the switching time of a micro-switch. Most micro-switches do not undergo uniform displacements, nor can these displacements be considered small, and the behaviour of a micro-switch is fundamentally transient.

To date, the most notable attempts to tackle the problem of reduced-order modelling (ROM) of squeeze-film damping with large, non-uniform displacements have been made by Younis *et al.* [5-7], Mehner *et al.* [4], Yang *et al.* [8], Hung and Senturia [9] and Rewienski and White [10]. In [5-7], the authors solve the linearized Reynolds equation for a displacement of the beam around an operating point. In [4], the authors use a modal projection method to calculate modal frequency-dependent damping and stiffening coefficients close to a determined operating point. To extend this approach to large displacements, Mehner [4] gives an analytical expression of these coefficients as a function of mechanical modal coordinates established by fitting of simulation data for different initial deformations. These approaches are all based on several steady-state sinusoidal calculations [5-7] or simulations [4], which increase the time for setting up the reduced-order model. The most general approaches may well be those developed in [8-10]. These approaches, although very general, have a high computational cost (because of the nonlinear /multiphysics /transient simulation they require) and their accuracy depends, to some degree, on the choice of the training trajectory.

## II. CONSTRUCTION OF THE REDUCED-ORDER MODEL

We work on the variable $p = P - P_0$, supposing $p \ll P_0$. (1) has then the following form:

$$\nabla\left(\frac{G^3}{12\mu}\nabla p\right) = \frac{\partial}{\partial t}\left(G\left(1+\frac{p}{P_0}\right)\right) \quad (2)$$

The first reduction step is based on modal projection of (2) which is first transformed via a change of variable on $p$ (3).

$$p = \varphi G^{(-3/2)} \quad (3)$$

The aim of this change of variable is to obtain a spatial operator on $\varphi$ for which the Laplacian eigenmodes are more relevant than for the operator in (2), conserving its self-adjoint property thus guarantying convergence of the solution. The reduced-order model may be written as:

$$\frac{d}{dt}(\mathbf{A}(\mathbf{x})\mathbf{s} - \mathbf{f}(\mathbf{x})) = \mathbf{H}(\mathbf{x})\mathbf{s} \quad (4)$$

with

$$f_l = 2P_0 \iint_\Omega \varphi_l G^{-1/2} d\Omega \quad (5)$$

$$H_{kl} = -\frac{P_0}{12\mu}\left(\lambda_k^2 \delta_{kl} + \iint_\Omega \frac{\Delta G^{3/2}}{G^{3/2}}\varphi_k \varphi_l d\Omega\right) \quad (6)$$

and

$$A_{kl} = \iint_\Omega G^{-2}\varphi_k \varphi_l d\Omega \quad (7)$$

where $\mathbf{x}$ and $\mathbf{s}$ are respectively the vectors of the mechanical modal coordinates of the moving structure, and the modified squeeze coordinates. For a structure under electrostatic actuation, one may write the full coupled model as:

$$\frac{d}{dt}\left(\begin{bmatrix} \mathbf{I} & 0 & 0 \\ 0 & \mathbf{M} & 0 \\ 0 & 0 & \mathbf{A}(\mathbf{x}) \end{bmatrix}\begin{bmatrix} \mathbf{x} \\ \dot{\mathbf{x}} \\ \mathbf{s} \end{bmatrix} - \begin{bmatrix} 0 \\ 0 \\ \mathbf{f}(\mathbf{x}) \end{bmatrix}\right) = \begin{bmatrix} 0 & \mathbf{I} & 0 \\ -\mathbf{K}(\mathbf{x}) & 0 & \mathbf{B}(\mathbf{x}) \\ 0 & 0 & \mathbf{H}(\mathbf{x}) \end{bmatrix}\begin{bmatrix} \mathbf{x} \\ \dot{\mathbf{x}} \\ \mathbf{s} \end{bmatrix} + \begin{bmatrix} 0 \\ \mathbf{p}_e(\mathbf{x}) \\ 0 \end{bmatrix} \quad (8)$$

where $\mathbf{p}_e(\mathbf{x})$ is the electrostatic force. The Euler-Bernoulli equation is used to model the mechanical part. Large displacement effects can be taken into account. Let us write (8) in a more compact way:

$$\frac{d}{dt}(\mathbf{g}(\mathbf{z}(t)) = \mathbf{f}(\mathbf{z}(t)) + \mathbf{B}(\mathbf{z}(t))u(t) \quad (9)$$

where $\mathbf{z}$ is the state vector including all modal coordinates.

## III. PIECEWISE LINEARIZATION BASED ON AN INPUT TRAJECTORY

The cost of evaluating the terms $\mathbf{g}(\mathbf{z}(t))$, $\mathbf{f}(\mathbf{z}(t))$, and $\mathbf{B}(\mathbf{z}(t))$ can be reduced using the piecewise linear approach described in [10]. The fact that the coefficients in (3) only depend on the mechanical modal coordinates reduces the cost of construction of the piecewise linear model, which has the following structure:

$$\frac{d}{dt}\left[\sum_{i=0}^{s-1} w_i(\mathbf{z})(\mathbf{g}(\mathbf{z}_i) + \mathbf{JG}_i(\mathbf{z}-\mathbf{z}_i))\right] = \sum_{i=0}^{s-1} w_i(\mathbf{z})(\mathbf{f}(\mathbf{z}_i) + \mathbf{JF}_i(\mathbf{z}-\mathbf{z}_i)) + \mathbf{F}_{elec}(\mathbf{z}) \quad (10)$$

where $\mathbf{JG}_i$ and $\mathbf{JF}_i$ are respectively the jacobians of the functions $\mathbf{g}(\mathbf{z})$ and $\mathbf{f}(\mathbf{z})$ at the linearization points $\mathbf{z}_i$ and $\mathbf{F}_{elec}(\mathbf{z})$ corresponds to the electrostatic force. Two problems arise from the piecewise linearization: the choice of the linearization points and the weighting procedure. We choose the linearization points from a simulation of the fully nonlinear reduced model (8). A new linearization point is chosen when a point is "far enough" from the already chosen points. The state variables must be normalized to calculate the distances in state-space. The weighting procedure is the one described in [14]. We work on the example of a microswitch also treated in [8-10]. We use the piecewise linear reduced order model to determine the switching time of the device for a step voltage between 9 and 10.5 V at atmospheric ambient pressure. The 21 linearization points are chosen along a 9.5V input training trajectory. The time integration scheme is not exactly implicit because the weights are calculated from the last step point. Fig.1 shows the response to a 10V input using a linear model, the modal projection model, and the piecewise linearized model. Fig.2 shows the experimental and simulated data presented in [9] and results of our piecewise linear reduced model of order 6 for the switching time.

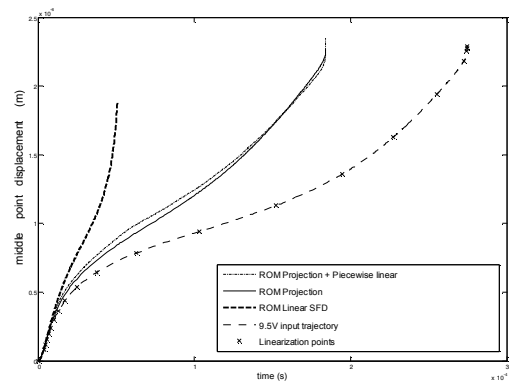

Fig. 1- Middle point displacement for the 9.5 V step voltage training trajectory, a 10 V full reduced model simulation, piecewise linearized model, and linear model..

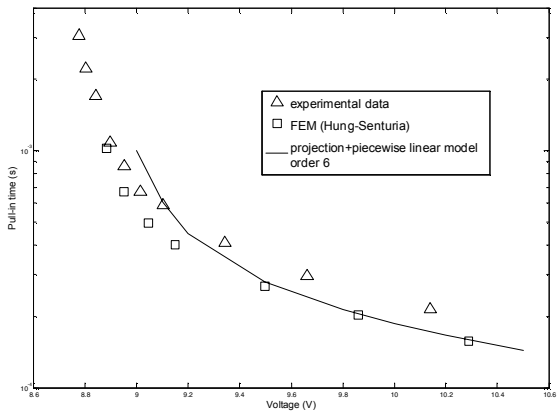

Fig. 2- Pull-in time (s) versus applied voltage (V) for $P_0 = 1.013 \times 10^5$ Pa. Comparison of the experimental and simulated results presented in [16] to the simulated results obtained with our reduced-order model. The chosen squeeze modes correspond to $k_x = 0, 2$ and $k_y = 1, 3$.

## IV. TRAJECTORY-INDEPENDENT REDUCED ORDER MODEL

We can notice that the nonlinear terms in (8) depend only on the mechanical modal coordinate and that the model is linear with respect to the modified squeeze coordinates. It can be written as:

$$\frac{d}{dt}(\mathbf{G}(\mathbf{x})\mathbf{z} - \mathbf{FP}(\mathbf{x})) = \mathbf{B}(\mathbf{x})\mathbf{z} + \mathbf{PE}(\mathbf{x}) \quad (11)$$

The piecewise linearization described in section II doesn't take advantage of this structure. This second piecewise linearized model is based on the linearization of the terms **G**, **FP** and **B**. As they only depend on the on mechanical coordinate, there is no need for a training trajectory. It is sufficient to discretize in an appropriate way the space corresponding to the mechanical coordinates. This is a great advantage as the resulting model does not depend on the relevance of a training trajectory, as opposed to the model presented in section II. On the other hand the resulting equation is nonlinear which increases the resolution cost.

| Number of linearization points | Middle point displacement error (in %) | Pull-in time error (in %) |
|---|---|---|
| 10 | 21 | 7 |
| 14 | 5 | 2.5 |
| 19 | 1 | 2 |
| 30 | 0.5 | 0.1 |

**Table 1** Middle point displacement error and pull-in time error for different number of linearization points compared to the full reduced model.

This equation is integrated in time using a fully implicit Euler scheme. At each step the nonlinear equation is solved using the Matlab "fsolve" function. Table 1 shows results for an input step voltage of 9.1 V. This model yields better displacement error and correct results are obtained for a minimum number of linearization points of around 15 which is less than the model obtained using the approach described in [10].

## V. CONCLUSION

We have presented a piecewise linear model of squeeze-film damping for flexible structure and large displacements with the restriction of small pressure variations. The model is based on a modal projection method and is then piecewise linearized using two different methods. The advantages and drawbacks of each method were pointed out. The first method [10] is very general whereas the second takes advantage of a specificity of the equation structure. This last method appears to be more efficient, although the final system that must be solved remains nonlinear. The pull-in time of a microswitch was determined and compared to experimental and other simulation data.